\documentclass[12pt]{article}
\usepackage{epsfig}
\usepackage{amssymb}
\usepackage{amsmath}
\usepackage{amsfonts}
\usepackage{graphicx}
\usepackage{mathrsfs}
\usepackage[dvips]{color}
\usepackage{multirow}

\newcommand{\R}{\mathbb{R}}
\newcommand{\C}{\mathbb{C}}

\newcommand{\fn}{{\,\mathfrak{n}\,}}

\newcommand{\fz}{\mathfrak{z}}

\newcommand{\bA}{\mathbf{A}}
\newcommand{\bB}{\mathbf{B}}
\newcommand{\bC}{\mathbf{C}}

\newcommand{\bH}{\mathbf{H}}

\newcommand{\bK}{\mathbf{K}}
\newcommand{\bM}{\mathbf{M}}

\newcommand{\bS}{\mathbf{S}}

\newcommand{\bU}{\mathbf{U}}

\newcommand{\cK}{\mathcal{K}}

\newcommand{\cO}{\mathcal{O}}

\newcommand{\be}{\begin{equation}}
\newcommand{\ee}{\end{equation}}
\newcommand{\bea}{\begin{eqnarray}}
\newcommand{\eea}{\end{eqnarray}}
\newcommand{\nn}{\nonumber}

\newcommand{\ed}{\end{document}}

\newcommand{\bi}{\begin{itemize}}
\newcommand{\ei}{\end{itemize}}

\newcommand{\bce}{\begin{center}}
\newcommand{\ece}{\end{center}}

\newcommand{\sH}{\mathscr{H}}

\newcommand{\sR}{\mathscr{R}}

\newcommand{\sT}{\mathscr{T}}

\newcommand{\RE}{{\rm Re}}

\newcommand{\lb}{\big[\hspace{-3pt}\big[}
\newcommand{\rb}{\big]\hspace{-3pt}\big]}
\newcommand{\Lsigma}{{\mbox{\large$\sigma$}}}

\begin{document}

\title{Composition of Transfer Matrices for Potentials with Overlapping Support}

\author{Farhang Loran\thanks{E-mail address:
loran@cc.iut.ac.ir}~ and Ali~Mostafazadeh\thanks{E-mail address:
amostafazadeh@ku.edu.tr}\\[6pt]
$^*$Department of Physics, Isfahan University of Technology,\\ Isfahan, 84156-83111, Iran\\[6pt]
$^\dagger$Departments of Mathematics and Physics, Ko\c{c} University,\\ 34450 Sar{\i}yer,
Istanbul, Turkey}

\date{ }
\maketitle

\begin{abstract}

For a pair of real or complex scattering potentials $v_j:\R\to\C$ ($j=1,2$) with support $I_j$ and transfer matrix $\bM_j$, the transfer matrix of $v_1+v_2$ is given by the product $\bM_2\bM_1$ provided that $I_1$ lies to the left of $I_2$. We explore the prospects of generalizing this composition rule for the cases that $I_1$ and $I_2$ have a small intersection. In particular, we show that if $I_1$ and $I_2$ intersect in a finite closed interval of length $\ell$ in which both the potentials are analytic, then the lowest order correction to the above composition rule is proportional to $\ell^5$. This correction is of the order of $\ell^3$, if $v_1$ and $v_2$ are respectively analytic throughout this interval except at $x=\ell$ and $x=0$. We use these results to explore the superposition of a pair of unidirectionally invisible potentials with overlapping support.
\vspace{2mm}

\noindent PACS numbers: 03.65.Nk\vspace{2mm}

\noindent Keywords: Scattering, transfer matrix, complex scattering potential, unidirectional invisibility
\end{abstract}

\section{Introduction}

Transfer matrices have numerous applications in a variety of scattering problems in physics and engineering \cite{razavy,sanchez}. This is mainly because of their composition property that allows for the determination of the scattering properties of a complicated system from the contributions of its simpler constituents. This is most simply described in the standard one-dimensional potential scattering \cite{muga}.

Consider a possibly complex-valued scattering potential $v:\R\to\C$ with an asymptotic decay rate such that \cite{faddeev-67}
    \be
    \int_{-\infty}^\infty (1+|x|)|v(x)|dx<\infty,
    \label{condi-math}
    \ee
and let $k$ be a positive real (wave)number. Then every solution of the Schr\"odinger equation
    \be
    -\psi''(x)+v(x)\psi(x)=k^2\psi(x),~~~~~~~~x\in\R,
    \label{sch-eq}
    \ee
satisfies
    \be
    \psi(x)\to A_\pm e^{ikx}+B_\pm e^{-ikx}\hspace{.5cm} {\rm as}\hspace{.5cm}x\to\pm\infty,
    \label{psi-asym}
    \ee
where $A_\pm $ and $B_\pm $ are possibly $k$-dependent complex coefficients \cite{muga}. The transfer matrix $\bM$ of the potential $v$ is a $k$-dependent  $2\times 2$ matrix that fulfills the relation \cite{razavy}
    \be
    \left[\begin{array}{c} A_+ \\ B_+ \end{array}\right]=
    \bM \left[\begin{array}{c} A_- \\ B_-\end{array}\right].
    \label{M-def}
    \ee
Similarly to the $S$-matrix, $\bM$ encodes the scattering properties of the potential $v$. Recalling that scattering solutions of (\ref{sch-eq}) are given in terms of the left/right reflection and transmission amplitudes, $R^{l/r}$ and $T$, according to
    \begin{align}
	&\psi_k^l(x)=\left\{
	\begin{array}{ccc}
	e^{ikx}+R^l e^{-ikx} & {\rm for} & x\to-\infty,\\
	T e^{ikx}& {\rm for} & x\to\infty,
	\end{array}\right.&&
	\psi_k^r(x)=\left\{ \begin{array}{ccc}
	T e^{-ikx} & {\rm for} & x\to-\infty,\\
	e^{-ikx}+R^r e^{ikx}& {\rm for} & x\to\infty,
	\end{array}\right.
	\label{scatter}
	\end{align}
and using (\ref{psi-asym}), (\ref{M-def}), and (\ref{scatter}), we can express the entries $M_{ij}$ of $\bM$ as \cite{prl-2009}
    \be
	\begin{aligned}
	&M_{11}=T- R^l R^r/T,~~~~~~ && M_{12}=R^r/T,\\
	&M_{21}=- R^l/T, && M_{22}=1/T.
	\end{aligned}
	\label{M-RT}
	\ee
These, in particular, imply that $\det \bM =1$. We also note that (\ref{condi-math}) is a sufficient condition for the (global) existence of the Jost solutions, $\psi_{k+}=\psi_k^l/T$ and $\psi_{k-}=\psi_k^r/T$, of (\ref{sch-eq}), \cite{kemp}.

\begin{figure}
	\begin{center}
	\includegraphics[scale=.4]{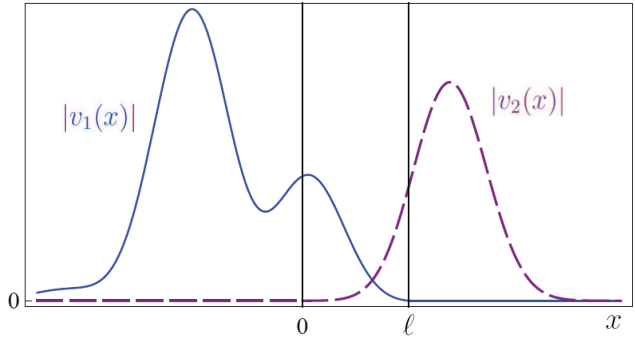}
	\caption{(Color online) Graphs of $|v_1(x)|$ (solid blue curve) and $|v_2(x)|$ (dashed purple curve) for a pair of potentials $v_i$ whose support $I_i$ satisfy $I_1\preccurlyeq I_2$. In particular, $I_1\cap I_2=[0,\ell]$ for some $\ell>0$.}
	\label{fig1}
	\end{center}
	\end{figure}

Now, suppose that $v$ can be written as the sum of a pair of scattering potentials $v_j:\R\to\C$ ($j=1,2$) with the same asymptotic decay property as $v$, such that the support\footnote{The support of a function $f:\R\to\C$ is the smallest closed interval outside which $f$ vanishes.} of $v_1$ lies to the left of that of $v_2$. Using $I_j$ to label the support of $v_j$, we express this condition by `$I_1\prec I_2$'.\footnote{`$I_1\prec I_2$' means that for every $x_1\in I_1$ and $x_2\in I_2$, we have $x_1\leq x_2$.} Under this assumption, we can relate $\bM$ to the transfer matrix $\bM_j$ of $v_j$ according to
    \be
    \bM=\bM_2\bM_1.
    \label{comp-prop}
    \ee
This is the celebrated `composition property' of the transfer matrix. It is also called the `group property', because transfer matrices belong to the matrix group $SL(2,\C)$ and (\ref{comp-prop}) involves the group multiplication for this group \cite{sanchez}. The primary aim of this article is to seek for a generalization of (\ref{comp-prop}) to the cases where $I_1\cap I_2$ is a finite interval. Without loss of generality, we can identify the latter with $[0,\ell]$, where $\ell$ is a real parameter signifying the length of $I_1\cap I_2$, and demand that for all $x_1\in I_1\setminus I_2$ and $x_2\in I_2$, $x_1<x_2$. We denote this relation by `$I_1\preccurlyeq I_2$.' Figure~\ref{fig1} provides a schematic description of this condition.    			
To summarize, we wish to generalize the composition property (\ref{comp-prop}) of the transfer matrix, which holds for $I_1\prec I_2$, to situations where $I_1\preccurlyeq I_2$.

The main tool that we employ in our treatment of this problem is a recent observation made in Ref.~\cite{ap-2014} which identifies the transfer matrix with the solution of an initial-value problem for a time-dependent Schr\"odinger equation. Let $[x_-,x_+]$ be any closed interval containing the support of $v$, $\tau_\pm=kx_\pm$, and $\bM(\tau,\tau_-)$ be the solution of
    \be
    i\partial_\tau\bM(\tau,\tau_-)=\sH(\tau)\bM(\tau,\tau_-),~~~~~~\bM(\tau_-,\tau_-)=\mathbf{1},
    \label{sch-eq-M}
    \ee
where $\tau\in[\tau_-,\tau_+]$,
    \be
    \sH(\tau):=w(\tau)\cK(\tau),~~~~~~
    \cK(\tau):=\left[\begin{array}{cc}
    1 & e^{-2i\tau}\\
    -e^{2i\tau} & -1\end{array}\right],~~~~~~~w(\tau):=\frac{v(\tau/k)}{2k^2},
    \label{H=}
    \ee
and $\mathbf{1}$ is the $2\times 2$ identity matrix. Then, $\bM=\bM(\tau_+,\tau_-)$. In other words, we can express $\bM$ as
    \be
    \bM=\sT \exp\left\{-i\int_{\tau_-}^{\tau_+}\sH(\tau)d\tau\right\},
    \label{time-order}
    \ee
where $\sT$ is the `time-ordering' operation. Notice also that for all $\tau_1,\tau_2\in[\tau_-,\tau_+]$ with $\tau_1<\tau_2$, the matrix $\bM(\tau_2,\tau_1)$ is the transfer matrix of the truncated potential:
    \be
    v_{\tau_1,\tau_2}(x):=\theta(kx-\tau_1)\theta(\tau_2-kx)v(x)=
    \left\{\begin{array}{ccc}
    v(x)&{\rm for}& kx\in[\tau_1,\tau_2],\\
    0 &{\rm for}& kx\notin[\tau_1,\tau_2],
    \end{array}\right.
    \label{truncated}
    \ee
where $\theta$ stands for the Heaviside step function; $\theta(\tau)=0$ if $\tau<0$, and $\theta(\tau)=1$ if $\tau\geq 0$. In other words,
    \be
    \bM(\tau_2,\tau_1)=\sT \exp\left\{-i\int_{\tau_1}^{\tau_2}\sH(\tau)d\tau\right\}.
    \label{time-order-2}
    \ee

A direct consequence of (\ref{sch-eq-M}) and (\ref{H=}) is that $\bM(\tau)^{-1\,T}$ satisfies
    \be
    i\partial_\tau \bM(\tau,\tau_-)^{-1\,T}=-\sH(\tau)^T \bM(\tau,\tau_-)^{-1\,T},~~~~~~~~~~
    \bM(\tau_-,\tau_-)^{-1\,T}=\mathbf{1},
    \label{sch-eq-inv}
    \ee
where the superscript `$T$' stands for the transpose of the corresponding matrix. More generally, we have
    \be
    \bM(\tau_2,\tau_1)^{-1\,T}=\sT\exp\left\{i\int_{\tau_1}^{\tau_2} \sH(\tau)^T d\tau\right\}.
    \label{time-order-3}
    \ee

\section{Generalized Composition Rule}

Consider the case where $I_1\preccurlyeq I_2$ and $I_1\cap I_2=[0,\ell]$ (See Fig.~\ref{fig1}.) Because $v=v_1+v_2$, we can take $[\tau_-,\tau_+]=I_1\cup I_2$. Moreover, let
    \begin{align}
    & v_-(x):=\theta(-x)v_1(x), && v_0(x):=\theta(x)\theta(\ell-x)v(x),
    && v_+(x):=\theta(x-\ell)v_2(x),
    \label{vs=}
    \end{align}
and $I_{\pm,0}$ and $\bM_{\pm,0}$ denote the support and transfer matrix of $v_{\pm,0}$, respectively. Then, $I_-\subseteq [\tau_-,0]$, $I_0\subseteq [0,\ell]$, and $I_+\subseteq [0,\tau_+]$. This shows that $I_-\prec I_0\prec I_+$. We also have $v_-+v_0+v_+=v_1+v_2=v$. Therefore, we can use the composition property of the transfer matrices, namely Eq.~(\ref{comp-prop}), to establish
    \be
    \bM=\bM_+\bM_0\bM_-.
    \label{M=3M}
    \ee

Next, we introduce the analog of $\bM(\tau_1,\tau_2)$ for the potentials $v_j$ with $j=1,2$; let $\bM_j(\tau_1,\tau_2)$  be given by replacing $v$ with $v_j$ on the right-hand side of (\ref{time-order-2}). Then (\ref{vs=}) implies
    \begin{align}
    &\bM_-=\bM_1(0,\tau_-), && \bM_0=\bM(\ell,0), && \bM_+=\bM_2(\tau_+,\ell).
    \label{Ms=Ms}
    \end{align}
Substituting these in (\ref{M=3M}) and using the fact that
    \begin{align*}
    &\bM_1=\bM_1(\ell,\tau_-)=\bM_1(\ell,0)\bM_1(0,\tau_-), &&
    \bM_2=\bM_2(\tau_+,0)=\bM_2(\tau_+,\ell)\bM_2(\ell,0),
    \end{align*}
we obtain
    \be
    \bM=\bM_2\,\bS(\epsilon)\,\bM_1,
    \label{general-comp}
    \ee
where $\epsilon:=k\ell$ and
    \be
    \bS(\epsilon):=\bM_2(\epsilon,0)^{-1}\bM(\epsilon,0)\bM_1(\epsilon,0)^{-1}.
    \label{S=}
    \ee
Introducing
    \be
    w_j(\tau):=\frac{v_j(\tau/k)}{2k^2},~~~~~~~~\sH_j(\tau):=w_j(\tau)\cK(\tau),
    \label{wj-Hj=}
    \ee
and using (\ref{time-order-3}), we can express $\bS(\epsilon)$ in the form
    \be
    \bS(\epsilon)=\mbox{$\left[\sT\exp\left\{i\int_0^\epsilon\sH_2(\tau)^T d\tau\right\}\right]^{\!T}\!
                \left[\sT\exp\left\{-i\int_0^\epsilon \sH(\tau) d\tau\right\}\right]\!
                    \left[\sT\exp\left\{\int_0^\epsilon\sH_1(\tau)^T d\tau\right\}\right]^T$}.
    \label{S=2}
    \ee
Notice that according to (\ref{H=}) and (\ref{wj-Hj=}), $\sH(\tau)=\sH_1(\tau)+\sH_2(\tau)$.

The matrix $\bS(\epsilon)$ contains all the scattering features of the potential $v$ stemming from the fact that the support of $v_1$ and $v_2$ overlap in the interval $[0,\ell]$; for $\ell=0$, Eq.~(\ref{S=2}) gives $\bS(0)=\mathbf{1}$. In the remainder of this section, we study the behavior of $\bS(\epsilon)$ for `small' $\epsilon$ by examining its expansion in powers of $\epsilon$;
    \begin{align}
    &\bS(\epsilon)=\mathbf{1}+\sum_{n=1}^\infty \frac{\bS_0^{(n)}\,\epsilon^n}{n!},
    \label{expand}\\
    & \bS_0^{(n)}:=\bS^{(n)}(0),
    ~~~~~~~~~\bS^{(n)}(\epsilon):=\partial^{\,n}_\epsilon\bS(\epsilon).
    \nn
    \end{align}

In order to compute the coefficient matrices $\bS_0^{(n)}$, we apply the above scheme for the truncated potentials $v_{j\tau}(x):=\theta(\tau-kx)v_j(x)$ where $\tau\in[0,k\ell]$. This amounts to changing $\epsilon$ in (\ref{S=}) into an independent parameter $\tau$;
    \be
    \bS(\tau):=\bM_2(\tau,0)^{-1}\bM(\tau,0)\bM_1(\tau,0)^{-1}.
    \label{S=tau}
    \ee
We can evaluate $\bS_0^{(n)}$ by repeatedly differentiating this equation with respect to $\tau$ and setting $\tau=0$.

Taking the derivative of both sides of (\ref{S=tau}) an using this equation together with (\ref{sch-eq-M}) and (\ref{sch-eq-inv}) in the resulting expression, we find that $\bS(\tau)$ is the solution of the initial-value problem:
    \bea
    &&i\dot\bS(\tau)=\lb\sH_1(\tau),\bS(\tau)\rb_\tau,
    \label{sch-eq-4}\\
    &&\bS(0)=\mathbf{1}.
    \label{S-zero}
    \eea
Here an overdot denotes differentiation with respect to $\tau$, and for any pair of $2\times 2$ matrix-valued functions $\bA$ and $\bB$, we have introduced
    \begin{align}
    &\lb\bA(\tau),\bB(\tau)\rb_\tau:=\widetilde\bA(\tau)\bB(\tau) -\bB(\tau) \bA(\tau) ,
    ~~~~~~~\widetilde\bA(\tau):=\bM_2(\tau)^{-1}\bA(\tau)\bM_2(\tau),
    \label{M2=4a}\\
    &\bM_2(\tau):=\bM_2(\tau,0)=\sT\exp\left\{-i\int_0^\tau\sH_2(\tau')d\tau'\right\}.
    \label{M2=4b}
    \end{align}
Observe that for $\tau=0$ the $\tau$-dependent bracket $\lb\cdot,\cdot\rb_\tau$ reduces to the usual commutator; \linebreak $\lb\bA(0),\bB(0)\rb_0 = [\bA(0),\bB(0)]$. Furthermore, the following useful identity holds.
    \be
    \partial_\tau \lb\bA(\tau),\bB(\tau)\rb_\tau =\lb\dot\bA(\tau),\bB(\tau)\rb_\tau+
    \lb\bA(\tau),\dot\bB(\tau)\rb_\tau-i\big[\widetilde\bA(\tau),\widetilde\sH_2(\tau)\big]\bB(\tau).
    \label{main-id}
    \ee

Equations~(\ref{sch-eq-4}) -- (\ref{M2=4b}) imply that if the potentials $v_j$ are analytic functions in $[0,\ell]$, then the series (\ref{expand}) converges. Because $v_1(x)=0$ for $x>\ell$ and $v_2(x)=0$ for $x<0$, this condition requires that $\lim_{x\to\ell^-}v_1(x)=v_1(\ell)=0$ and $\lim_{x\to 0^+}v_2(x)=v_2(0)=0$, which are rather restrictive. It is not difficult to see that the convergence of (\ref{expand}) can also be established for the cases where $v_1$ and $v_2$ are analytic in $(0,\ell)$ but are respectively allowed to be discontinuous at $x=\ell$ and $x=0$. In the following we focus our attention to potentials $v_j$ with these properties, and, without loss of generality, identify the value of $v_1$ (respectively $v_2$) at $x=\ell$ (respectively $x=0$) with its left limit as $x\to\ell$ (respectively right limit as $x\to 0$.) The same applies for the first and higher order derivatives of $v_1$ (respectively $v_2$) at $x=\ell$ (respectively $x=0$); we identify them with the corresponding left (respectively right) derivatives.

Next, we list some of the properties of the matrix $\cK(\tau)$ that we will use in the calculation of $\bS_0^{(n)}$. They follow directly from the definition of $\cK(\tau)$, i.e., (\ref{H=}).
    \begin{align}
    & \cK_0:=\cK(0)=i\Lsigma_2+\Lsigma_3,
    &&\dot\cK_0:=\dot\cK(0)=-2i\Lsigma_1,
    && \ddot\cK_0:=\ddot\cK(0)=4(\Lsigma_3-\cK_0),
    \label{algebra-01}\\
    & \cK(\tau)=e^{-i\tau\Lsigma_3}\cK_0\, e^{i\tau\Lsigma_3},
    && \dot\cK(\tau)=e^{-i\tau\Lsigma_3}\dot\cK_0\, e^{i\tau\Lsigma_3},
    && \ddot\cK(\tau)=e^{-i\tau\Lsigma_3}\ddot\cK_0\, e^{i\tau\Lsigma_3},
    \label{algebra-02}\\
    & \cK(\tau)^2=\cK_0^2=\mathbf{0},
    &&\dot\cK(\tau)^2=\dot\cK_0^2=-4 \mathbf{1},
    &&[\Lsigma_3,\cK(\tau)]=i\dot\cK(\tau),
    \label{algebra-03}\\
    & [\cK(\tau),\dot\cK(\tau)]=-4i\,\cK(\tau),
    && [\cK(\tau),\ddot\cK(\tau)]=-4i\,\dot\cK(\tau),
    && [\Lsigma_3,\dot\cK(\tau)]=i\ddot\cK(\tau),
    \label{algebra-11}
    \end{align}
where $\mathbf{0}$ and $\mathbf{1}$ respectively denote the $2\times 2$ null and identity matrices, and $\Lsigma_i$, with $i=1,2,3$, stand for the Pauli matrices.\footnote{According to (\ref{algebra-03}) and (\ref{algebra-11}), for all $\tau\in\R$, the matrices $\bK_1:=-\dot\cK(\tau)/4$, $\bK_2:=\ddot\cK(\tau)/8$, and $\bK_3:=\Lsigma_3/2$ are generators of the algebra $su(1,1)$; they satisfy
$[\bK_1,\bK_2]=i\bK_3$, $[\bK_2,\bK_3]=i\bK_1$, and $[\bK_3,\bK_1]=-i\bK_2$.}

Now, we are in a position to compute $\bS_0^{(n)}$. We begin by noting that in view of (\ref{sch-eq-4}) and (\ref{S-zero}),
    \be
    \bS^{(1)}_0:=\dot\bS(0)=\mathbf{0}.
    \label{S-one}
    \ee
Next, we use (\ref{wj-Hj=}), (\ref{M2=4a}), and the first equation in (\ref{algebra-03}), to establish
    \be
    \widetilde\sH_i(\tau)\widetilde\sH_j(\tau)= \sH_i(\tau)\sH_j(\tau) =\mathbf{0}.
    \label{id2}
    \ee
If we differentiate both sides of (\ref{sch-eq-4}) and use Eqs.~(\ref{main-id}) and (\ref{id2}) to simplify the result, we arrive at
    \be
    i \ddot\bS(\tau)=\lb \dot\sH_1(\tau),\bS(\tau)\rb_\tau+\lb\sH_1(\tau),\dot\bS(\tau)\rb_\tau.
    \label{2nd-der-eq}
    \ee
Setting $\tau=0$ in this equation and employing (\ref{S-zero}) and (\ref{S-one}) yield
    \be
    \bS^{(2)}_0:=\ddot\bS(0)=\mathbf{0}.
    \label{S-two}
    \ee

We can similarly carry out the calculation of $\bS^{(3)}(\tau)$ and $\bS^{(4)}(\tau)$. Taking the derivative of (\ref{2nd-der-eq}) and using (\ref{main-id}) and (\ref{id2}) give
    \be
    i\bS^{(3)}(\tau)=\lb\ddot\sH_1(\tau),\bS(\tau)\rb_\tau+2\lb\dot\sH_1(\tau),\dot\bS(\tau)\rb_\tau+
    \lb\sH_1(\tau),\ddot\bS(\tau)\rb_\tau-i[\widetilde{\dot\sH_1}(\tau),\widetilde\sH_2(\tau)]
    \bS(\tau).
    \label{3rd-der-eq}
    \ee
In light of (\ref{wj-Hj=}), (\ref{S-zero}), (\ref{S-one}), (\ref{S-two}), and the first equation in (\ref{algebra-11}), Eq.~(\ref{3rd-der-eq}) implies
	\be
	\bS^{(3)}_0:=\bS^{(3)}(0)=-4i\,w_1(0)w_2(0)\,\cK_0.
	\label{S-three}
	\ee
To determine $\bS^{(4)}(\tau)$, we make use of (\ref{main-id}) to differentiate (\ref{3rd-der-eq}). This gives
    \bea
    i\bS^{(4)}(\tau)&=&
    \lb \sH_1^{(3)}(\tau),\bS(\tau)\rb_\tau+3\lb\ddot\sH_1(\tau),\dot\bS(\tau)\rb_\tau+
    3\lb\dot\sH_1(\tau),\ddot\bS(\tau)\rb_\tau+\lb\sH_1(\tau),\bS^{(3)}(\tau)\rb_\tau
    \nn\\
    & &
    -i\Big\{ [\widetilde{\ddot\sH_1}(\tau),\widetilde\sH_2(\tau)]\bS(\tau)+
    2[\widetilde{\dot\sH_1}(\tau),\widetilde\sH_2(\tau)]\dot\bS(\tau)
    +\frac{d}{d\tau}\Big([\widetilde{\dot\sH_1}(\tau),\widetilde\sH_2(\tau)]\bS(\tau)\Big)
    \Big\},~~~~~~
    \label{4th-der-eq}
    \eea
where we have also benefitted from (\ref{id2}). In view of (\ref{wj-Hj=}), (\ref{S-zero}), (\ref{algebra-11}), (\ref{S-one}), (\ref{id2}), (\ref{S-two}), and (\ref{S-three}), Eq.~(\ref{4th-der-eq}) implies
    \be
    \bS^{(4)}_0:=\bS^{(4)}(0)=-16\,w_1(0)w_2(0)\,\Lsigma_1-4i
    \big[w_1(0)\dot w_2(0)+3\,\dot w_1(0)w_2(0)\big]\cK_0.
    \label{S-four}
    \ee

Substituting (\ref{S-one}), (\ref{S-two}), (\ref{S-three}), and (\ref{S-four}) in (\ref{expand}), we find the following more explicit expression for $\bS(\epsilon)$.
    \bea
    \bS(\epsilon)&=&\mathbf{1}-\frac{2 i}{3}\, w_1(0)\, w_2(0)\,\cK_0\,\epsilon^3\nn\\
    &&-\frac{1}{6}\Big\{ 4 w_1(0)\, w_2(0)\,\Lsigma_1+i\big[w_1(0)\, \dot w_2(0)+3\dot w_1(0)\, w_2(0)\big]\cK_0\Big\}\,\epsilon^4+
    {\cal O}(\epsilon^5)\nn\\
    &=&
    \mathbf{1}-\frac{2 i}{3}\, w_1(\epsilon)\, w_2(0)\,\cK_0\,\epsilon^3\nn\\
    &&-\frac{1}{6}\Big\{ 4 w_1(\epsilon)\, w_2(0)\,\Lsigma_1+
    i\big[w_1(\epsilon)\, \dot w_2(0)-\dot w_1(\epsilon)\, w_2(0)\big]\cK_0\Big\}\,\epsilon^4+
    {\cal O}(\epsilon^5)\nn\\
     &=&
    \mathbf{1}-\frac{i}{3!\,k}\, v_1(\ell)\, v_2(0)\,\cK_0\,\ell^3\nn\\
    &&-\frac{1}{4!\, k}\Big\{4k\, v_1(\ell)\, v_2(0)\,\Lsigma_1+
    i\big[v_1(\ell)\, v'_2(0)-v'_1(\ell)\, v_2(0)\big]\cK_0\Big\}\,\ell^4+
    {\cal O}(\ell^5),
    \label{S-leading-order-3n}
    \eea
where $\cO(\epsilon^d)$ stands for terms of order $d$ and higher in powers of $\epsilon$, and
we have used the Taylor series expansion of $w_1(\tau)$ about $\epsilon$ for $\tau=0$, namely
    \be
    w_1(0)=w_1(\epsilon)+\sum_{n=1}^\infty \frac{(-1)^n}{n!}\:w_1^{(n)}(\epsilon)\epsilon^n=
    w_1(\epsilon)-\dot w_1(\epsilon)\epsilon+\cO(\epsilon^2),
    \label{w1-Taylor}
    \ee
to establish the second equality.

Equation~(\ref{S-leading-order-3n}) implies that whenever both $v_1(\ell)$ and $v_2(0)$ are nonzero,
the leading order correction to the standard composition rule for transfer matrices is of the order of $\ell^3$. This correspond to situations where $v_1$ and $v_2$ are respectively discontinuous at $x=\ell$ and $x=0$, because $I_1\subseteq (-\infty,\ell]$ and
$I_2\subseteq [0,\infty)$. If $v_1$ is continuous at $x=\ell$, we have $v_1(\ell)=0$, and (\ref{S-leading-order-3n}) reduces to
    \be
    \bS(\epsilon)=\mathbf{1}+\frac{i}{4!\, k}\, v'_1(\ell)\, v_2(0)\,\cK_0 \,\ell^4+
    {\cal O}(\ell^5).
    \label{S-leading-order-4n}
    \ee
Therefore, the leading order correction terms is at least of order $\ell^4$. The same holds for cases where $v_2$ is continuous at $x=0$. In this case,
    \be
    \bS(\epsilon)=\mathbf{1}-\frac{i}{4!\, k}\, v_1(\ell)\, v'_2(0)\,\cK_0 \,\ell^4+
    {\cal O}(\ell^5).
    \label{S-leading-order-4nn}
    \ee

For situation where $v_1$ and $v_2$ are respectively continuous at $x=\ell$ and $x=0$, the leading order correction term is at least of the order of $\ell^5$. In order to obtain the explicit form of this term, we must calculate $\bS^{(n)}_0$ for $n\geq 5$. This is a tedious task. In the appendix we outline a different scheme to conduct this calculation. It gives the following improvement of (\ref{S-leading-order-3n}).
    \bea
    \bS(\epsilon)&=&\mathbf{1}-\frac{i}{3!\,k}\,v_1(\ell)v_2(0)\,\cK_0\,\ell^3\nn\\
    &&-\frac{1}{4!\,k}\,\Big\{4 k\, v_1(\ell)v_2(0)\Lsigma_1+
    i\big[v_1(\ell) v'_2(0)-v'_1(\ell)v_2(0)\big]\cK_0\Big\}\,\ell^4\nn\\
    && -\frac{1}{5!\,k}\,\Big\{
    k\big[6\,v_1(\ell)v'_2(0)-4v'_1(\ell)v_2(0)\big]\Lsigma_1
    +8k^2\,v_1(\ell)v_2(0)\Lsigma_2\nn\\
    &&~~~~~~~~~+i\big[-4k^2\,v_1(\ell)v_2(0)-v'_1(\ell)v'_2(0)
    +v_1(\ell)v''_2(0)+v''_1(\ell)v_2(0)\nn\\
    &&~~~~~~~~~~~~~~~+4\{v_1(\ell)v_2(0)^2+v_1(\ell)^2v_2(0)\}\big]
    \cK_0\Big\}\,\ell^5
    \nn\\
    &&-\frac{1}{6!\,k}\,\Big\{
    \big[-32k^3\,v_1(\ell)v_2(0)-6k\,v'_1(\ell)v'_2(0)
    +4k\{2 v_1(\ell)v''_2(0)+v''_1(\ell) v_2(0)\}\nn\\
    &&~~~~~~~~~~~
    +16k\, v_1(\ell)v_2(0)\{2v_1(\ell)+v_2(0)\}\big]\Lsigma_1
    +4k^2\big[5\,v_1(\ell)v'_2(0)-2 v'_1(\ell) v_2(0)\big]\Lsigma_2\nn\\
    &&~~~~~~~~~~+i\big[4k^2\{v'_1(\ell)v_2(0)-v_1(\ell)v'_2(0)\}
    +v''_1(\ell)v'_2(0)-v'_1(\ell)v''_2(0)\nn\\
    &&~~~~~~~~~~~~~~+v_1(\ell)v_2^{(3)}(0)-v_1^{(3)}(\ell)v_2(0)
    +4\{v_1(\ell)^2v'_2(0)-v'_1(\ell)v_2(0)^2\}\nn\\
    &&~~~~~~~~~~~~~~-14\,v_1(\ell)v_2(0)\{v'_1(\ell)-v'_2(0)\}
    \big]\cK_0\Big\}\,\ell^6 +{\cal O}(\ell^7).
    \label{S-leading-order-6-new}
    \eea
In particular, whenever $v_1(\ell)=v_2(0)=0$, we have
    \begin{align}
    \bS(\epsilon)=&
    \mathbf{1}+\frac{i}{5! k}\,v'_1(\ell) v'_2(0)\,\cK_0\,\ell^5\nn\\
    &+\frac{1}{6!\,k}\left\{6k\,v'_1(\ell)v'_2(0)\,\Lsigma_1+
    i\big[v'_1(\ell) v''_2(0)-v''_1(\ell)v'_2(0)\big]\cK_0\right\}\ell^6 +{\cal O}(\ell^7).
    \label{S-leading-order-6-ell}
    \end{align}

\section{Application to Unidirectional Invisibility}

Consider the potential
    \be
    V(x):=\left\{\begin{array}{cc}
    \fz\, e^{iK x} & {\rm for}~0\leq x\leq L,\\
    0 & {\rm otherwise},\end{array}\right.
    \label{unidir}
    \ee
where $\fz,K$ and $L$ are real parameters. This is the first-known example of a complex scattering potential that displays unidirectional reflectionlessness and invisibility \cite{invisible-1,invisible-2,pra-2014a,jpa-2014b,pra-2014b} (for sufficiently small $\fz$.) In particular, if
    \be
    k=\frac{K}{2}=\frac{2\pi m}{L},
    \label{condi-undir}
    \ee
for some positive integer $m$, then the reflection and transmission amplitudes of (\ref{unidir}) take the form \cite{pra-2014a}:
    \begin{align}
    & R^l=\cO(\fz^3), &&
    R^r=m\left[\sR^{(1)}\fz+\sR^{(2)}\fz^2\right]+\cO(\fz^3),
    && T=1+m\,\sT^{(2)}\fz^2+\cO(\fz^3),
    \label{RRT=1}
    \end{align}
where
    \begin{align*}
    &\sR^{(1)}:=\frac{-iL^2}{4\pi m^2}=-\frac{4\pi i}{K^2},
    &&\sR^{(2)}:=\frac{iL^4}{32\pi^3 m^4}=\frac{8\pi i}{K^4},
    &&\sT^{(2)}:=\frac{\sR^{(2)}}{4}.
    \end{align*}
According to (\ref{RRT=1}), the potential (\ref{unidir}) is left-reflectionless if we can neglect $\cO(\fz^3)$, and left-invisible if we can neglect $\cO(\fz^2)$. Following \cite{pra-2014a} we call these properties perturbative unidirectional reflectionlessness and invisibility, respectively.

Let $V_{K,m}(x)$ denote the potential (\ref{unidir}) subject to the condition,
	\[L=L_m:=\frac{4\pi m}{K},\]
and consider the situation where
    \begin{align}
    &v_1(x):=V_{K,1}(x+L_1-\ell),  &&v_2(x):=V_{K,1}(x), && v(x):=v_1(x)+v_2(x).
    \label{VVV=}
    \end{align}
Clearly, the support $I_j$ of $v_j$ are given by
    \begin{align*}
    &I_1=[-L_1+\ell,\ell], && I_2=[0,L_1],
    \end{align*}
so that $I_1\cap I_2=[0,\ell]$. It is also easy to see that for $\ell=0$, $v(x)=V_{K,2}(x+L_1)$. Because $|R^{l/r}|$ are invariant under space translations \cite{pra-2014b}, $v(x)$ is also perturbatively reflectionless from the right for $\ell=0$. In the following, we use the results of the preceding section to explore the extent to which this property is violated for $\ell\neq 0$.

First, we compute the transfer matrices $\bM_j$ for $v_j$ up to and including terms of order $\fz^2$ for $k=K/2$. We can easily do this by realizing that under a translation, $x\to x-d$, the transfer matrix $\bM$ of any given potential transforms according to $\bM\to e^{-ikd\,\Lsigma_3}\bM\, e^{ikd\,\Lsigma_3}$. Equivalently, we have $R^l\to e^{2i kd}R^l$, $R^r\to e^{-2i kd}R^r$, and $T\to T$, \cite{pra-2014b}. In view of (\ref{M-RT}), (\ref{condi-undir}), (\ref{RRT=1}) and (\ref{VVV=}), these imply
    \begin{align}
    &\bM_1=\left[\begin{array}{cc}
    1+\sT^{(2)}\fz^2 & e^{-i K \ell}\big(\sR^{(1)}\fz+\sR^{(2)}\fz^2\big)\\
    0 & 1-\sT^{(2)}\fz^2\end{array}\right]+\cO(\fz^3),
    \label{M1=zz}\\
    &\bM_2=\left[\begin{array}{cc}
    1+\sT^{(2)}\fz^2 & \sR^{(1)}\fz+\sR^{(2)}\fz^2 \\
    0 & 1-\sT^{(2)}\fz^2\end{array}\right]+\cO(\fz^3).
    \label{M2=zz}
    \end{align}

Next, we compute $\bS(\epsilon)$. It is easy to see that because $v_1(\ell)=v_2(0)=\fz$ and
$v'_1(\ell)=v'_2(0)=iK\fz$, Eq.~(\ref{S-leading-order-3n}) gives
    \begin{align}
    \bS(\epsilon)=&
    \mathbf{1}-\frac{i\fz^2}{3K}\,\cK(0)\,\ell^3-\frac{\fz^2}{6}\,\Lsigma_1\,\ell^4 +{\cal O}(\ell^5).
    \label{unidir-S}
    \end{align}
Here we have also made use of (\ref{condi-undir}).

Substituting (\ref{M1=zz}) -- (\ref{unidir-S}) in (\ref{general-comp}), we find that for $k=K/2$ the transfer matrix of the potential $v$ is given by
    \bea
    \bM&=&\bM_2\bM_1+\bS(\epsilon)-\mathbf{1}+\cO(\fz^3)\nn\\
    &=&\left[\begin{array}{cc}
    1+2\big(\sT^{(2)}+\tilde\sT^{(2)}\big)\fz^2 & (1+e^{-iK\ell})\left(\sR^{(1)}\fz+
    \sR^{(2)}\fz^2\right)-\tilde\sR^{(2)}_+\fz^2\\
    -\tilde\sR^{(2)}_-\fz^2 & 1-2\big(\sT^{(2)}+
    \tilde\sT^{(2)}\big)\fz^2\end{array}\right]+\cO(\fz^3,\ell^5),~~~
    \label{M=MSM=zz}
    \eea
where
    \begin{align}
    &\tilde\sT^{(2)}:=-\frac{i\ell^3}{6K}, &&
    \tilde\sR^{(2)}_\pm:=\frac{\pm 2i\ell^3+ K\ell^4}{6K}.
    \end{align}
With the help of (\ref{M-RT}) and (\ref{M=MSM=zz}), we can compute the reflection and transmission amplitudes of this potential for $k=K/2$. These have the form
    \bea
    R^l&=&\tilde\sR^{(2)}_-\fz^2+\cO(\fz^3,\ell^5)=
    \frac{8\fz^2}{3K^4}\left(-i\,\epsilon^3+\epsilon^4\right)+\cO(\fz^3,\epsilon^5),
    \label{RL=zz1}
        \\
    R^r&=&\left(1+e^{-i K\ell}\right)\big(\sR^{(1)}\fz+\sR^{(2)}\fz^2\big)-\tilde\sR^{(2)}_+\fz^2
    +\cO(\fz^3,\ell^5)\nn\\
    &=&4 i\left(1+e^{-2i\,\epsilon}\right)\left(-\frac{\fz}{K^2}+\frac{2\pi\fz^2}{K^4}\right)
    -\frac{8\fz^2}{3 K^4}\left(i\,\epsilon^3+\epsilon^4\right)+\cO(\fz^3,\epsilon^5),
    \label{RR=zz1}
        \\
    T&=&1+2\big(\sT^{(2)}+\tilde\sT^{(2)}\big)\fz^2+\cO(\fz^3,\ell^5)=
    1+\frac{2i\fz^2}{K^4}\left(2\pi-\frac{4\,\epsilon^3}{3}\right)+\cO(\fz^3,\epsilon^5).
    \label{T=zz1}
    \eea
As seen from these formulas, small overlaps between the support of $v_1$ and $v_2$ do not affect the unidirectional invisibility of their sum up to order $\fz^2$, but it slightly violates their unidirectional reflectionless up to order $\fz^3$. This is, in a sense, a sign of robustness of the phenomenon of unidirectional invisibility.

In typical optical applications, $v(x)=k^2[1-\fn(x)^2]$, where $\fn(x)$ is the refractive index of the medium \cite{SS-review}. This suggests that $|\fz|/K^2$ is of the same order of magnitude as $|\fn(x)^2-1|$. Because for non-exotic material the real part of $\fn$ is much larger than its imaginary part, the condition $|\fz|/K^2\ll 1$ corresponds to optical media with $\RE(\fn)\approx 1$. To realize optical potentials of the form (\ref{unidir}), one must dope a host medium with $\RE(\fn)\approx 1$ and try to generate the needed loss/gain profile by properly pumping it. This would necessarily involve various errors including those related to the location and size of the pumped region. Our results show that these errors do not obstruct the unidirectional invisibility and reflectionlessness of the sample.

\section{Concluding Remarks}

The composition or group property of transfer matrices is the main reason for their popularity and usefulness. This property applies whenever one wishes to determine the scattering features of a potential that is the sum of two or more constituent potentials with mutually disjoint support.
In the present article we have explored a way of extending this property to a pair of constituent potentials $v_j$ with overlapping support.

For situations where the support of $v_j$ intersect in a finite interval $[0,\ell]$, the standard composition rule, i.e., $\bM=\bM_2\bM_1$, generalizes to $\bM=\bM_2\bS\,\bM_1$, where $\bS$ depends on $\ell$ as well as the behavior of $v_j$ in $[0,\ell]$. Assuming that $v_i$ are analytic functions in $(0,\ell)$, we can compute $\bS$ in a power series in $\epsilon=k\ell$, where $k$ is the wavenumber. We have shown that if $v_1$ and $v_2$ are respectively analytic functions at $x=\ell$ and $x=0$, so that $v_1(\ell)=v_2(0)=0$, then the leading order term in the expansion of $\bS-\mathbf{1}$ is of order $\epsilon^5$. Otherwise this term is of order $\epsilon^3$ or $\epsilon^4$.

Our results reveals another interesting fact regarding the scaling properties of the power series expansion of $\bS$. The coefficient matrices $\bS_0^{(n)}$ for $n\geq 0$, which determine this expansion, scale at least quadratically under the scaling of the potential: $v_j\to \alpha v_j$, where $\alpha$ is a constant. This shows that the first-order perturbation theory is not capable of detecting the correction factor $\bS-\textbf{1}$, i.e., the standard composition rule for transfer matrices, namely  (\ref{comp-prop}), applies whenever the first-order perturbation theory is proven reliable.

\subsection*{Acknowledgments} This project was initiated during F.~Loran's visit to Ko\c{c} University in  November 17-29, 2014. We are indebted to the Turkish Academy of Sciences (T\"UBA) for providing the financial support which made this visit possible. This work has been supported by the Scientific and Technological Research Council of Turkey (T\"UB\.{I}TAK) in the framework of the project no: 112T951 and by T\"UBA.

\section*{Appendix: Power series for $\bS(\epsilon)$}

In this appendix we outline a method of evaluating the coefficient matrices $\bS_0^{(n)}$ appearing in the $\epsilon$-series expansion (\ref{expand}) of $\bS(\epsilon)$. Because
$\bS(\tau):=\bM_2(\tau,0)^{-1}\bM(\tau,0)\bM_1(\tau,0)^{-1}$, we should examine the $\epsilon$-series expansion for $\bM(\epsilon,0)$. It turns out that it is easier to work with $\bU(\tau):=e^{i\,\tau\,\Lsigma_3}\,\bM(\tau,0)$, which satisfy
    \be
    \bS(\epsilon)=\bU_2^{-1}(\epsilon)\,\bU(\epsilon)\,\bU^{-1}_1(\epsilon)\,
    e^{i\,\epsilon\,\Lsigma_3}.
    \label{che}
    \ee
Therefore, we first study the $\epsilon$-series expansion of $\bU(\epsilon)$ and $\bU(\epsilon)^{-1}$.

Because \begin{align}
    &\cK(\tau)=e^{-i\,\tau\,\Lsigma_3}\,\cK_0\,e^{i\,\tau\,\Lsigma_3},
    &&\cK_0=\Lsigma_3+i\,\Lsigma_2=\left[\begin{array}{cc}
    1 & 1\\ -1 & -1\end{array}\right],
    \label{K-tau}
    \end{align}
$\bU(\tau)$ fulfils the matrix Schr\"odinger equation,
    \begin{align}
    &i\,\dot \bU(\tau)=\bH(\tau)\bU(\tau),
    \label{orig}
    \end{align}
for the Hamiltonian
    \be
    \bH(\tau):=w(\tau)\,\cK_0-\Lsigma_3.
    \label{H=app}
    \ee
We also have $\bU(0)=\mathbf{1}$. Therefore, $\bU(\epsilon)={\cal T}\exp\left\{-i\int_0^\epsilon d\tau\,\bH(\tau)\right\}$.

The Hamiltonian $\bH(\tau)$ has been introduced in \cite{pra-2014a} and studied extensively in \cite{jpa-2014a} (See also \cite{jpa-2014b}.) It enjoys the following useful property.
    \be
    \bH(\tau)^T=\Lsigma_3\,\bH(\tau)\,\Lsigma_3.
    \label{H-transpose}
    \ee
This relation is a consequence of the fact that $\cK_0^T=\Lsigma_3\,\cK_0\,\Lsigma_3$. As noted in \cite{pra-2014a}, $\cK(\tau)$ is a $\Lsigma_3$-pseudo-Hermitian matrix \cite{p123}, i.e., $\cK(\tau)^\dagger=\Lsigma_3\,\cK(\tau)\,\Lsigma_3^{-1}$.

In view of (\ref{orig}) and (\ref{H=app}), we can easily verify that
     \bea
     \bU(\epsilon)&=&
     \left({\bf 1}\, \exp\left\{\left[\overleftarrow{\partial_\tau}
     -i\,\bH(\tau)\right]\epsilon\right\}\right|_{\tau=0}=\sum_{n=0}\frac{\epsilon^n}{n!}\,
     \mathbf{C}_n,
     \label{U-der}
     \eea
where
     \be
     \mathbf{C}_n^T:=\left.\left\{\left[\overrightarrow{\partial_\tau }-i\,\bH(\tau)^T\right]^n
     {\bf 1}\right\}\right|_{\tau=0}=
     \left.\left\{\left[\overrightarrow{\partial_\tau }-i\,[w(\tau)\,\cK(0)^T-\Lsigma_3]\right]^n
     {\bf 1}\right\}\right|_{\tau=0}.
     \label{Cn}
     \ee
Similarly, we can show that
    \be
    \bU(\epsilon)^{-1}=\sum_{n=0}\frac{\epsilon^n}{n!} \, \check\bC_n=\Lsigma_2\, \bU(\epsilon)^T\,\Lsigma_2,
    \label{U-inv=}
    \ee
where
    \be
    \check\bC_n:=\left.\left\{\left[\overrightarrow{\partial_\tau }+
    i\,\bH(\tau)\right]^n{\bf 1}\right\}\right|_{\tau=0}=\Lsigma_2\bC_n^T\,\Lsigma_2.
    \label{C-check}
    \ee
Here the last equality follows from (\ref{H-transpose}) and (\ref{Cn}).

In order to determine $\bC_n$, we introduce the projection matrices:
     \begin{align}
     &\Delta:=\left[\begin{array}{cc}1&1\\0&0\end{array}\right],
     &&\Gamma:=\left[\begin{array}{cc}1&-1\\0&0\end{array}\right],
     \label{D-G}
     \end{align}
and note that, according to \eqref{orig}, $\bU(\tau)$ satisfies
     \begin{align}
     &i\,\Delta \dot \bU(\tau)=-\Gamma \bU(\tau),
     &&
     i\,\Gamma \dot \bU(\tau)=f(\tau)\, \Delta \bU(\tau),
     \label{zq}
     \end{align}
where $f(\tau):=2\,w(\tau)-1$. It is not difficult to see that $\Gamma \bU(\tau)$ and $\Delta  \bU(\tau)$ determine $\bU(\tau)$ uniquely. Indeed, every $2\times 2$ matrix $\bA$ satisfies
    \be
    \frac{1}{2}\left[ (\Delta\bA+\Gamma\bA)+\Lsigma_1(\Delta\bA-\Gamma\bA)\right]=\bA.
    \label{identity-A}
    \ee
We, therefore, proceed to explore $\Delta\bU(\tau)$ and $\Gamma\bU(\tau)$.

Clearly, $\Delta\bU(\tau)$ has the form
    \be
    \Delta\bU(\tau)=\left[\begin{array}{cc}u_+(\tau)&u_-(\tau)\\0&0\end{array}\right],
    \label{delta-U=1}
    \ee
for a pair of complex-valued functions $u_\pm$, and in view of Eq.~(\ref{zq}) and $\bU(0)=\mathbf{1}$, it satisfies
    \begin{align}
    &\partial_\tau^2\Delta\bU(\tau)=f(\tau)\,\Delta\bU(\tau),
    \label{b1}\\[6pt]
    &\Delta\bU(0)=\Delta,~~~~~~~~~~~
    \partial_\tau\Delta\bU(\tau)\Big|_{\tau=0}=i\,\Gamma.
    \end{align}
These in turn imply that $u_\pm(\tau)$ are the solutions of the second order linear homogenous equation:
    \be
    \ddot u(\tau)-f(\tau)\,u(\tau)=0,
    \label{eq2-xpm}
    \ee
that fulfil the initial conditions:
    \be
    u_\pm(0)=1,\hspace{1cm}\dot u_\pm(0)=\pm \,i.
    \label{ini-condi-xpm}
    \ee

Next, we obtain a power series solution of (\ref{eq2-xpm}). To this end we express derivatives of $u$ in the form
  \be
  u^{(n)}(\tau)=p_n(\tau)\,u(\tau)+q_n(\tau)\,\dot u(\tau),
  \label{zzq12}
  \ee
where $p_n$ and $q_n$ are a pair of auxiliary functions. In light of (\ref{eq2-xpm}) and (\ref{zzq12}), they satisfy
    \begin{align}
    &\left[
    \begin{array}{c}p_{n+1}(\tau)\\q_{n+1}(\tau)\end{array}\right]
    ={\cal D}(\tau)\left[\begin{array}{c}p_{n}(\tau)\\q_{n}(\tau)\end{array}\right],
    && \left[\begin{array}{c}p_{1}(\tau)\\q_{1}(\tau)\end{array}\right]=
    \left[\begin{array}{c}0\\1\end{array}\right],
    \end{align}
where $n\geq 1$ and ${\cal D}(\tau):=\left[\begin{array}{cc}
    \partial_\tau&f(\tau)\\1&\partial_\tau\end{array}\right]$. In particular, $\left[
    \begin{array}{c}p_{n+1}(\tau)\\q_{n+1}(\tau)\end{array}\right]
    ={\cal D}^n(\tau)\left[\begin{array}{c}0\\1\end{array}\right]$.
Using this relation in (\ref{zzq12}) and employing (\ref{ini-condi-xpm}), we find
    \be
    u_\pm(\epsilon)=1+\sum_{n=0}\frac{\epsilon^{n+1}\,d^\pm_{n}}{(n+1)!},
    \label{xpm-exp}
    \ee
where for all $n\geq 0$,
    \be
    d^\pm_n:=\left.\left\{\left[\begin{array}{cc}1&\pm \,i\end{array}\right] {\cal D}(\tau) ^{n}\left[\begin{array}{c}0\\1\end{array}\right]\right\}\right|_{\tau=0}.
    \label{d}
    \ee

Now, we substitute (\ref{xpm-exp}) in (\ref{delta-U=1}) to obtain $\Delta\bU$. This together with
(\ref{zq}) and (\ref{identity-A}) yield
    \be
    \bU(\epsilon)=1 +\sum_{n=0}\frac{\epsilon^{n+1}}{2(n+1)!}\left[  \begin{array}{cc}
    d^+_n-i\,d^+_{n+1}&d^-_n-i\,d^-_{n+1}\\
    d^+_n+i\,d^+_{n+1}&d^-_n+i\,d^-_{n+1}
    \end{array}\right].
    \label{U-exp=sp}
    \ee
Comparing this relation to (\ref{U-der}), we find that the coefficient matrices $\bC_n$ of Eq.\eqref{Cn} have the form
     \be
     \bC_{n+1}=\frac{1}{2}\left[  \begin{array}{cc}
     d^+_n-i\,d^+_{n+1}&d^-_n-i\,d^-_{n+1}\\d^+_n+i\,d^+_{n+1}&d^-_n+i\,d^-_{n+1}
     \end{array}\right].
     \label{Cn=expl}
     \ee
This relation together with (\ref{U-inv=})  and (\ref{C-check}) allow us to compute the $\epsilon$-series expansion of  $\bU^{-1}_1(\epsilon)$ and $\bU^{-1}_2(\epsilon)$. Using these and (\ref{U-exp=sp}) in (\ref{che}) yields the desired expression for $\bS(\epsilon)$. In the following we apply this method to compute $\bS^{(n)}_0$ for $n\leq 6$.

First, we employ (\ref{d}) to compute
    \begin{align*}
    &d^\pm_0=\pm\,i,~~~~~~~~~~~~~~~~~~~~d^\pm_1=f,~~~~~~~~~~~~~~~~~~d^\pm_2=\dot f\pm\,i\,f,\\
    &d^\pm_3=\ddot f+f^2\pm\,2\,i\,\dot f,
    ~~~~~~~~~~~~~d^\pm_4=f^{(3)}+4\,f\,\dot f\pm\,i\,(3\,\ddot f+f^2),\\
    &d^\pm_5=f^{(4)}+4{\dot f}^2+7\,f\,\ddot f+f^3\,\pm\, i\,(4\,f^{(3)}+6\,f\,\dot f),\\
    &d^\pm_6=f^{(5)}+15\,\dot f\,\ddot f+11\,f\,f^{(3)}+9\,f^2\,\dot f \pm  i\,(5\,f^{(4)}+10\,{\dot f}^2+13\,f\,\ddot f+f^3).
    \end{align*}
Here and in what follows $f$  and its derivatives $f^{(n)}$ are evaluated at $\tau=0$. Using these equations in (\ref{Cn=expl}), we obtain
    \begin{align*}
    &\bC_1=i\Big[\Lsigma_3-\frac{1}{2}(f+1)\,\cK(0)\Big],~~~~~~~~~~~~
    \bC_2=f\mathbf{1}-\frac{i\,\dot f}{2}\,\cK(0),\\
    &\bC_3=\frac{1}{2}\,\dot f(3 \mathbf{1}-\Lsigma_1)
    -\frac{i}{2}\Big[(\ddot f+f^2)\,\cK(0)-f\cK(0)^T\Big],\\
    &\bC_4=(2\ddot f+f^2)\mathbf{1}-\ddot f\,\Lsigma_1+i\,\dot f\,\cK(0)^T
    -i\Big(2 f \dot f+\frac{1}{2}\,f^{(3)}\Big)\cK(0),\\
    &\begin{aligned}
    \bC_5=&\,\frac{5}{2}\Big(f^{(3)}+2 f \dot f\Big)\mathbf{1}
        -\Big(\frac{3}{2}\,f^{(3)}+f\,\dot f\Big)\,\Lsigma_1\\
        &+\frac{1}{2}\left(f^{(4)}+f^3+4\,{\dot f}^2+7\,f\,\ddot f+3\,\ddot f+f^2\right)\,\Lsigma_2\\
        &-\frac{i}{2}\left(f^{(4)}+f^3+4\,{\dot f}^2+7\,f\,\ddot f-3\,\ddot f-f^2\right)\,\Lsigma_3,
    \end{aligned}
    \end{align*}
    \begin{align*}
    &\begin{aligned}
    \bC_6&=\Big(3\,f^{(4)}+7\,{\dot f}^2+{10\,f\,\ddot f}+f^3\Big)\textbf{1}
    -\left(2\,f^{(4)}+3\,{\dot f}^2+3\,f\,\ddot f\right)\,\Lsigma_1\\
        &+\frac{1}{2}\left(f^{(5)}+15\,\dot f\,\ddot f+11\,f\,f^{(3)}+9\dot f\,f^2+4\,f^{(3)}+6\,f\,\dot f\right)\,\Lsigma_2\\
        &-\frac{i}{2}\left(f^{(5)}+15\,\dot f\,\ddot f+11\,f\,f^{(3)}+9\,\dot f\,f^2-4\,f^{(3)}-6\,f\,\dot f\right)\,\Lsigma_3.
    \end{aligned}
    \end{align*}
Substituting these in (\ref{U-der}) and (\ref{C-check}), and using (\ref{U-inv=}), we can determine  $\bU(\epsilon)$, $\bU_1(\epsilon)^{-1}$, and $\bU_2(\epsilon)^{-1}$. If we insert the resulting expressions in (\ref{che}) and compare the result with (\ref{expand}), we can compute $\bS^{(n)}_0$ for $n\leq 6$. The result is
	\begin{align}
	\bS^{(1)}_0=&~\bS^{(2)}_0=\mathbf{0},\quad\quad\quad\quad\quad\quad\quad\quad\quad
	\bS^{(3)}_0=-4i\,w_1(0)\,w_2(0)\,\cK(0),
	\label{S0123=}\\
	\bS^{(4)}_0=&
	-16\,w_1(0) w_2(0)\,\Lsigma_1-4i\Big[ w_1(0)\dot w_2(0)+3\,w_2(0)\dot w_1(0)\Big]\cK(0),
	\label{S04=}\\
	\bS^{(5)}_0=
	&-4\Big[16\,\dot w_1(0)w_2(0)+6\,w_1(0)\dot w_2(0)\Big]\,\Lsigma_1
	-32\,w_1(0)w_2(0)\,\Lsigma_2\nn\\
	 &~~~~~~-4i\,\Big[4\,w_1(0)w_2(0)\left(2\,w_1(0)+2\,w_2(0)-1\right)
	+w_1(0)\ddot w_2(0)\nn\\
	&~~~~~~~~~~~~~~+6\,\ddot w_1(0)w_2(0)+4\,\dot w_1(0)\dot w_2(0)\Big]\,\cK(0),
	\label{S05=}\\
	 \bS^{(6)}_0=&-8\Big[16\,w_1(0)w_2(0)\{2w_1(0)+w_2(0)-1\}+
	15\,\dot w_1(0)\dot w_2(0)\nn\\
	&~~~~~~+4\,w_1(0)\ddot w_2(0)+20\,\ddot w_1(0)w_2(0)\Big]\Lsigma_1
	-80\Big[w_1(0)\dot w_2(0)+2\,\dot w_1(0)w_2(0)\Big]\Lsigma_2\nn\\
	&-4i\Big[4 w_1(0)w_2(0)\{17\dot w_1(0)+7\,\dot w_2(0)\}
	+4w_1(0)\dot w_2(0)\{2\,w_1(0)-1\}\nn\\
	&~~~~~~~+20\,\dot w_1(0)w_2(0)\{2\,w_2(0)-1\}+5\dot w_1(0)\ddot w_2(0)+
	10\,\{\ddot w_1(0)\dot w_2(0)+w_1^{(3)}(0)w_2(0)\}\nn\\
	&~~~~~~~+w_1(0)w_2^{(3)}(0)\Big]\cK(0).
	\label{S06=}
	\end{align}

Next, we recall that we can assume, without loss of generality, that $v_1(\epsilon)$ coincides with its left limit at $x=\epsilon$. This allows for expanding $w_1(\tau)$ and its derivatives in Taylor series about $\tau=\epsilon$ for $\tau<\epsilon$. In particular, we have
    \begin{align*}
    &w_1(0)=w_1(\epsilon)-\dot w_1(\epsilon)\,\epsilon+\frac{1}{2}\,\ddot w_1(\epsilon)\,\epsilon^2
    -\frac{1}{6}w_1^{(3)}(\epsilon)\,\epsilon^3+{\cal O}(\epsilon^4),\\
    &\dot w_1(0)=\dot w_1(\epsilon)-\ddot w_1(\epsilon)\,\epsilon+
    \frac{1}{2}\, w_1^{(3)}(\epsilon)\,\epsilon^2+{\cal O}(\epsilon^3),\\
    &\ddot w_1(0)=\ddot w_1(\epsilon)-w_1^{(3)}(\epsilon)\,\epsilon+{\cal O}(\epsilon^2),~~~~~~~~~~
    w_1^{(3)}(0)=w_1^{(3)}(\epsilon)+{\cal O}(\epsilon).
    \end{align*}
Substituting these equations  in (\ref{S0123=}) -- (\ref{S06=}), inserting the result in (\ref{expand}), and noting that $\tau=kx$ and $w_j(\tau)=v_j(x)/2k^2$, we obtain
(\ref{S-leading-order-6-new}).

\ed